\documentclass[3p,times,twocolumn]{elsarticle}

\usepackage{ecrc}
\usepackage{hyperref}
\usepackage{amsmath}
\usepackage{xcolor}
\hypersetup{colorlinks=true,linkbordercolor=blue,linkcolor=blue, citecolor=blue,pdfborderstyle={/S/U/W 1}}
\usepackage{url}
\def\url#1{\href{#1}{{\tt #1}}}

\volume{00}

\firstpage{1}

\journalname{Nuclear and Particle Physics Proceedings}

\runauth{Fran\c{c}ois Arleo}

\jid{nppp}

\jnltitlelogo{Nuclear and Particle Physics Proceedings}

\usepackage{amssymb}
\usepackage[figuresright]{rotating}

\newcommand{\bc}{\begin{center}}
\newcommand{\ec}{\end{center}}
\newcommand{\be}{\begin{equation}}
\newcommand{\ee}{\end{equation}}
\newcommand{\bi}{\begin{itemize}}
\newcommand{\ei}{\end{itemize}}
\newcommand{\tf}{t_{_{\rm f}}}
\newcommand{\A}{\rmsf{A}}
\def\sqrts{{\sqrt{s}}}

\newcommand{\Z}{{\rm Z}}
\newcommand{\RpA}{R_{_{p\A}}}
\def\rmsf#1{{\rm \sf #1}}
\def\cO#1{{{\cal{O}}}\left(#1\right)}
\def\kt{{k_{_\perp}}} 
\def\jpsi{J/\psi}
\def\pt{p_{_\perp}}

\begin{document}

\begin{frontmatter}

\dochead{}

\title{Aspects of hard QCD processes in proton--nucleus collisions}

\author{Fran\c{c}ois Arleo}

\address{Laboratoire Leprince-Ringuet, \'Ecole polytechnique, CNRS/IN2P3,  Universit\'e Paris-Saclay,  91128, Palaiseau, France}

\begin{abstract}
Hard processes in high-energy proton--nucleus collisions are a powerful tool in order to investigate several importants aspects of QCD in a nuclear medium, such as nuclear shadowing, parton multiple scattering or medium-induced gluon radiation. I review in these proceedings recent progress in that field.
\end{abstract}


\end{frontmatter}

%%%%%%%%%%%%%%%%%%%%%%%%%%%%%%%%%%%%%%%%%
\section{Introduction}
\label{sec:intro}
%%%%%%%%%%%%%%%%%%%%%%%%%%%%%%%%%%%%%%%%%

Given the title of this conference, the utility of hard processes to probe QCD media -- whether nuclear matter or the quark-gluon plasma -- in nuclear collisions needs little introduction. First of all, their variety is appealing: measurements in heavy-ion collisions at RHIC and at LHC include electroweak processes (prompt photons, weak bosons), inclusive and heavy-flavour jets, as well as light and heavy hadrons. Moreover, most of these processes are well understood in QCD, meaning that perturbative calculations exist --~either fixed order, typically next-to-leading order (NLO) in the strong coupling constant $\alpha_s$, or including resummation~-- and accurately describe pp collision data; this is, however, less true when it comes to the production of hadrons, and in particular quarkonium production which remains far from being understood. Very much discussed at this conference, jet quenching and quarkonium suppression are among the most spectacular manifestations of quark-gluon plasma formation in heavy-ion collisions. Extracting the physical properties of the QCD plasma from these experimental observations remains however a delicate, albeit exciting, challenge.

Proton--nucleus collisions are (slightly) less complex than heavy-ion collisions, but no less interesting. In these reactions, the QCD medium under consideration, `cold' nuclear matter, is simpler than the quark-gluon plasma: it is static, with a known nuclear density profile (despite the fluctuating positions of the nucleons inside the nucleus). At the LHC, bulk observables in pPb collisions may point to the formation of a hot medium; its influence on hard processes, however, often appears limited (a possible exception being the production of excited quarkonia). Moreover, on the experimental side measurements are more easily performed due to the lower multiplicity underlying event. In short, the field of hard processes in pA collisions aims at the precision of the QCD studies performed in pp collisions while allowing for a quantitative study of nuclear medium effects in a controlled environment.

In these proceedings I will discuss several QCD phenomena expected to affect the rate of hard processes in pA collisions, either within QCD collinear factorization (Section~\ref{sec:lt}) or clearly beyond this framework (Section~\ref{sec:ht}). Section~\ref{sec:hardsoft} is devoted to the event activity dependence and the correlations between soft underlying event and hard process.

%%%%%%%%%%%%%%%%%%%%%%%%%%%%%%%%%%%%%%%%%
\section{Collinear factorization}
\label{sec:lt}
%%%%%%%%%%%%%%%%%%%%%%%%%%%%%%%%%%%%%%%%%

Let us first consider a generic hard process in pp collisions, characterized by a hard scale $Q$ much larger than a typical hadronic scale, $Q \gg \Lambda=\cO{1\ {\rm GeV}}$. (Think of the production of a large transverse momentum parton or a massive weak boson.) According to QCD collinear factorization, the production cross section can be written symbolically as
\be
\label{eq:xspp}
\sigma^{{\rm pp}} =  f_{i}^{p}(\mu)\ \otimes\ f_{j}^{p}(\mu)\ \otimes\ {\hat{\sigma}}_{\rm ij}(\mu, \mu^\prime)\ +\ 
\cO{\alpha_s^k}\ +\ 
\cO{\frac{\Lambda^n}{Q^n}}\ , \quad
\ee
in which long distance physics is encoded into parton distribution functions (PDF) of the incoming protons, $f^{p}$, while the partonic cross section, $\hat{\sigma}$, represents short distance physics\footnote{Factorization and renormalization scales are of order $\mu \sim \mu^\prime \sim Q$ to avoid the appearance of large logarithms $\ln(Q/\mu)$.}.
Collinear factorization exhibits strong predictive power: (i) PDF are non-perturbative but universal quantities, which can be probed either in deep inelastic scattering or in hadronic collisions, and (ii) the short distance scattering is computable (at least in principle, if not in practice) at any order in perturbation theory. One should also keep in mind that collinear factorization is an approximation. On top of neglecting higher-order terms in the perturbative expansion, process-dependent power corrections (so-called `higher-twist') of order\footnote{Typically $n=1$ or $n=2$ depending on the specific process.} $\cO{\Lambda^n/Q^n}$ might contribute significantly to the cross section when the scale $Q$ is not too large.

What about collinear factorization in pA collisions? Regarding the nucleus as any other hadron, the cross section Eq.~(\ref{eq:xspp}) could be written as
\be
\nonumber
\sigma^{{\rm pA}} =  f_{i}^{p}(\mu)\ \otimes\ f_{j}^{A}(\mu)\ \otimes\ {\hat{\sigma}}_{\rm ij}(\mu, \mu^\prime)\ +\ 
\cO{\alpha_s^k}\ +\ 
\cO{\frac{\Lambda_{_{\rm A}}^n}{Q^n}}\ , \quad
\ee
with $f^{A}$ now being the PDF of the nucleus. Note the new scale $\Lambda_{_{\rm A}}$ controlling the power corrections which could, in principle, increase with the target size. Consequently higher-twist processes are likely to be enhanced in pA collisions with respect to pp collisions.

An important question is what to expect for $f^{\rm A}$. Imagine a super dilute nucleus, in which nucleons are separated over macroscopic distances: the PDF of such a nucleus would simply be given by the incoherent sum over the proton and neutron PDF. The leading twist cross section (neglecting power suppressed corrections for the time being) would thus simply be expressed as $\sigma^{\rm pA} = Z\  \sigma^{\rm pp} + (A-Z)\  \sigma^{\rm pn} \simeq A\  \sigma_{\rm pp}$ (assuming $\sigma^{\rm pp} \simeq \sigma^{\rm pn}$ for QCD processes at high energy) making the nuclear production ratio, $\RpA  \equiv 1/A\ \sigma^{\rm pA}/\sigma^{\rm pp}$, normalized to unity. In practice, however, the typical distance between nucleons, say 1~fm, is much smaller than the  (Ioffe) length over which the hard process develops, $\ell_c = 1/(2 m x_2)$, $x_2$ being the target parton momentum fraction of the nucleon and $m$ the nucleon mass. When $\ell_c \gtrsim 1$~fm, equivalently at small $x_2 \lesssim 10^{-1}$, several nucleons in the target contribute coherently to the hard process, leading to the depletion of nuclear PDF (nPDF) ratios, $R_i\equiv f_i^A/Af_i^p<1$, known as shadowing. The precise determination of shadowing, and more generally nPDF at any value of $x$, is thus an important requirement in order to predict accurately the yields of hard processes in pPb collisions at the LHC.

As is the case for their proton counterparts, the nPDF cannot be computed from first principles (although their evolution, in either $x$ or $Q^2$, is perturbative) and need to be extracted from global fits to data. Over the last years various nPDF sets based on DGLAP evolution have been extracted at NLO accuracy and a first attempt (KA15) has been made recently at NNLO~\cite{Eskola:2009uj}. Due to the lack of data on nuclear targets the present nPDF sets still suffer from large uncertainties, especially at small $x$ and in the gluon sector; for the same reason it was shown that the present nPDF global fits suffer from a strong sensitivity on their parametrization at the input scale~\cite{Helenius:2016hcu}. Using reweighting techniques~\cite{Paukkunen:2014zia}, the present and future LHC pPb data can be used to significantly narrow the uncertainties of the nPDF sets currently available~\cite{Armesto:2015lrg}.
What are the best processes to constrain nuclear parton densities at the LHC? Some `ideal' requirements (not strictly necessary but which I consider preferable) are listed:
\begin{itemize}
\item[(i)] The scale $Q$ should be `large enough' compared to the saturation scale of the nucleus, $Q \gg Q_s$ (typically $Q_s \simeq1$--3~GeV at the LHC), in order to avoid the appearance of non-linear evolution effects not taken into account in the nPDF global fits; this would also suppress large power corrections entering pA cross sections. Note however that nPDF effects are expected to vanish at very large scales ($R_i^A(x, Q^2\to\infty)\to1$) because of QCD evolution; to illustrate this, the gluon nPDF ratio given by EPS09 at $x=10^{-3}$, $R_g^{\rm Pb}=0.84$ at $Q^2 = 10~{\rm GeV}^2$ while $R_g^{\rm Pb}=0.96$ at $Q^2 = 10^4~{\rm GeV}^2$. Therefore $Q$ should not be chosen too large in order to keep some sensitivity in the data;
\item[(ii)] Due to multiple scattering, the well-known modification of the $\pt$ spectrum of particles produced in pA collisions (the so-called `Cronin effect') is likely to spoil a clean extraction of nPDF. Such an effect nevertheless disappears for $\pt$-integrated cross sections or at $\pt \gg Q_s$. In other words, it may be safer \emph{not} to use $\pt$-differential cross sections at moderate $\pt$ values, say $\pt \lesssim 10$~GeV;
\item[(iii)] The production of color neutral hard probes should be preferred as they are insensitive to energy loss effects (discussed in Section~\ref{sec:ht}) which might affect dramatically the rate of other hard processes, such as light and heavy hadrons.
\end{itemize}
Based on the above discussion, the most promising candidates to probe nPDF might be the production of inclusive jets/dijets~\cite{Paukkunen:2014pha}, massive weak bosons~\cite{Paukkunen:2010qg}, and low-mass Drell-Yan pairs~\cite{Arleo:2015qiv}. Other interesting processes include prompt photon~\cite{Arleo:2007js} and top quark  production~\cite{dEnterria:2015mgr}.

High-precision measurements on dijet pseudo-rapidity distributions in pp and pPb collisions have been reported by the CMS experiment at this conference~\cite{CMS:2016kjd}. Data span a wide pseudo-rapidity coverage ($-2.0 < \eta_{\rm dijet} < 2.8$) in various $\pt$ bins ($25 \lesssim \pt \lesssim 200$~GeV), which allow for probing nPDF on a large range in $x$ and $Q^2$. 
While the magnitude of the dijet $\RpA$ ratio is well captured by most nPDF sets, it is interesting to note that none of the existing nPDF sets reproduce the rapidity trend for all $\pt$ bins. 
At negative $\eta_{\rm dijet}$, data seem to agree better with DSSZ which predict a small suppression in the large-$x$ `EMC region'; on the contrary, the suppression measured at forward pseudo-rapidity is better reproduced by EPS09 and nCTEQ15 sets, which shadowing at small $x$ is more pronounced than that of DSSZ.
Together with the measurements to be performed in pPb collisions at $\sqrt{s}=8.16$~TeV, dijets promise to bring stringent constraints on gluon  and valence quark nPDF in the intermediate $x$ range, $x \sim 10^{-2}$--$10^{-1}$.
%%%%%%%%%%%%%%%%%%%%%%%%%%%%%%%%%%%%%%%%%%
\begin{figure}[tbp]
\bc \includegraphics[width=7.4cm]{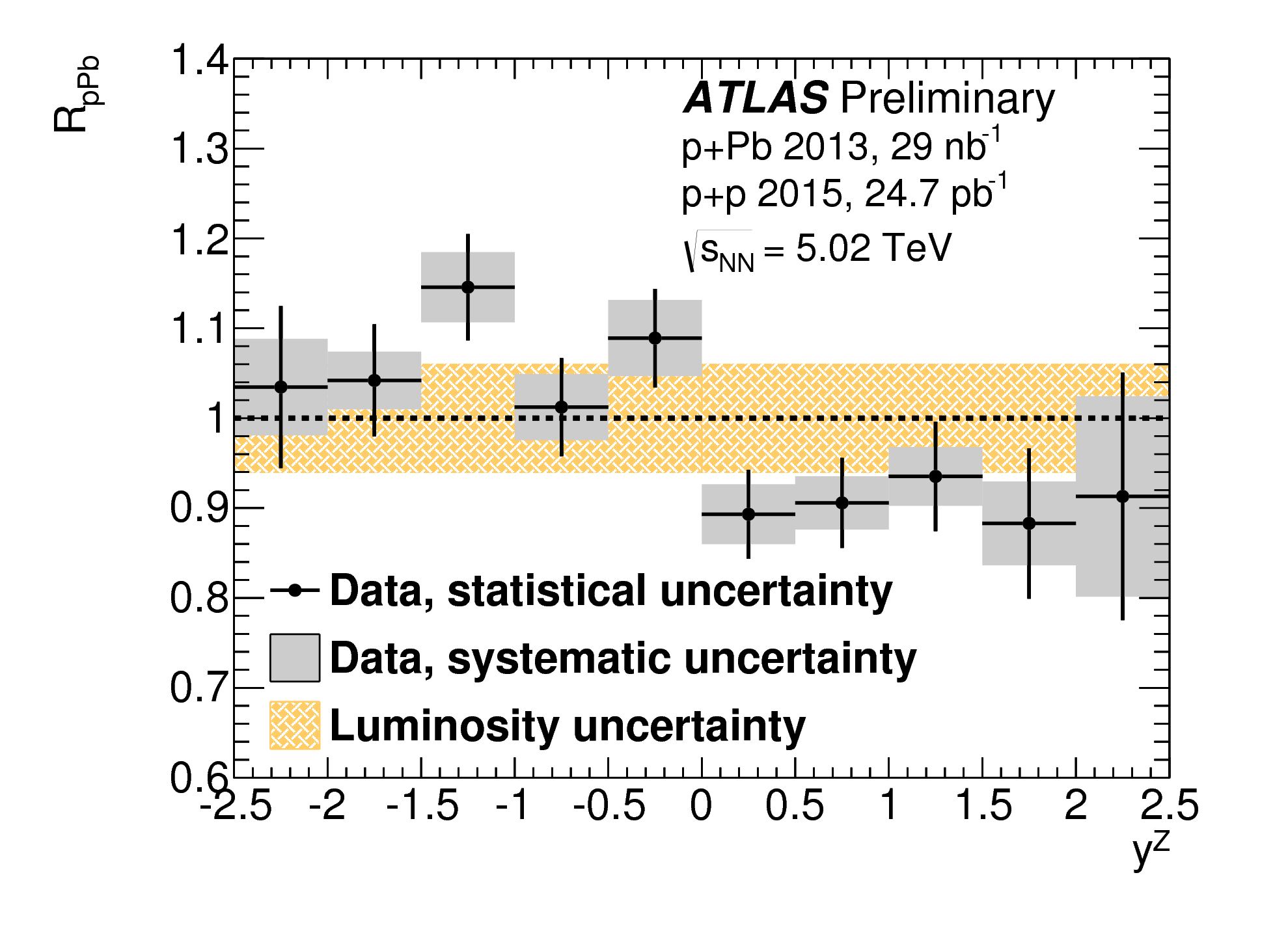}  \ec
\vspace{-.4cm}
\caption{$\RpA^{\Z}$ measured by ATLAS~\cite{ATLAS:2016ckg}.}
   \label{fig:z_atlas}
\vspace{-.2cm}
\end{figure}
%%%%%%%%%%%%%%%%%%%%%%%%%%%%%%%%%%%%%%%%%%

New data on weak boson production in pPb collisions have also been reported by ATLAS~\cite{ATLAS:2016ckg}, following earlier measurements by {ALICE}, ATLAS and CMS. As seen in Fig.~\ref{fig:z_atlas}, a slight depletion of Z boson is observed at forward rapidity (note however the systematic uncertainty from the luminosity determination), which could hint at sea quark shadowing even at this large scale, $Q^2\sim10^4$~GeV$^2$. Similarly, in the semileptonic W-decay channel, CMS measurements in pPb collisions are best reproduced by NLO calculations when nPDF corrections (as given by EPS09) are taken into account~\cite{Khachatryan:2015hha,Ru:2016ifu}. The rapidity distribution in the lepton charge asymmetry, $(\ell^+-\ell^-)/(\ell^++\ell^-)$, known to be sensitive to the $d/u$ PDF ratio (in a proton, or in a nucleus at  negative $\eta_{\ell}$~\cite{Arleo:2015dba}), differs slightly from the calculations at $\eta_{\ell}\lesssim-1$. This could be interpreted as a different nuclear modification of the up and down quark partonic density, $R_u^{\rm Pb} \ne R_d^{\rm Pb}$. Interestingly, a similar observation is reported at this conference by ATLAS, in some classes of event activity~\cite{ATLAS:2016ckg}. Data collected during the pPb run at $\sqrt{s}=8.16$~TeV in November 2016 should be able to clarify shortly the origin of a possible disagreement.

The large data sample collected during this successful pPb run should also make possible the measurement of Drell-Yan (DY) lepton pairs of low mass, $M_{_{\ell\ell}}\gtrsim 10$~GeV. The DY mechanism offers many advantages. It is a clean process whose production is known in QCD to a very good accuracy, at fixed order (NNLO) and including resummation~\cite{Catani:2009sm}. As discussed in Ref.~\cite{Arleo:2015qiv}, energy loss effects on DY are expected to be negligible, making this process an excellent probe of nuclear parton densities at a low scale ($Q\sim M_{_{\ell\ell}}$) where nPDF corrections are expected to be the largest. This is in sharp contrast to W/Z bosons, jets and photons which probe nPDF at a much larger scale. Drell-Yan measurements at large rapidity, accessible in the LHCb acceptance~\cite{Graziani:2145943}, are thus expected to give tight constraints on the degree of sea quark shadowing at small values of $x_2 \sim 10^{-5}$--$10^{-4}$~\cite{Arleo:2015qiv}. 

%%%%%%%%%%%%%%%%%%%%%%%%%%%%%%%%%%%%%%%%%
\section{Beyond collinear factorization}
\label{sec:ht}
%%%%%%%%%%%%%%%%%%%%%%%%%%%%%%%%%%%%%%%%%

So far I have discussed observables which could help the determination of the leading-twist nPDF, from a global fit to data based on collinear factorization. This framework, however, does not include the multiple scattering of the incoming parton while traversing the nucleus, which has two important consequences:
\begin{itemize}
\item[(i)] The incoming parton acquires additional transverse momentum, ${\langle \pt^2 \rangle}_{\rm pA} - {\langle \pt^2 \rangle}_{\rm pp} \simeq \hat{q} L = Q_s^2$, for a large nucleus, 
where $\hat{q}$ is the nuclear matter transport coefficient and $L$ the medium length. This transverse momentum broadening leads to Cronin effect, the distortion of $\pt$-spectra at $\pt \lesssim Q_s$;
\item[(ii)] In addition, multiple soft scattering induce the emission of soft gluons from the projectile parton. This leads to medium-induced energy loss which affects the rapidity distribution of hard processes.
\end{itemize}
%%%%%%%%%%%%%%%%%%%%%%%%%%%%%%%%%%%%%%%%%%
\begin{figure}[tbp]
\bc
\includegraphics[width=6.5cm]{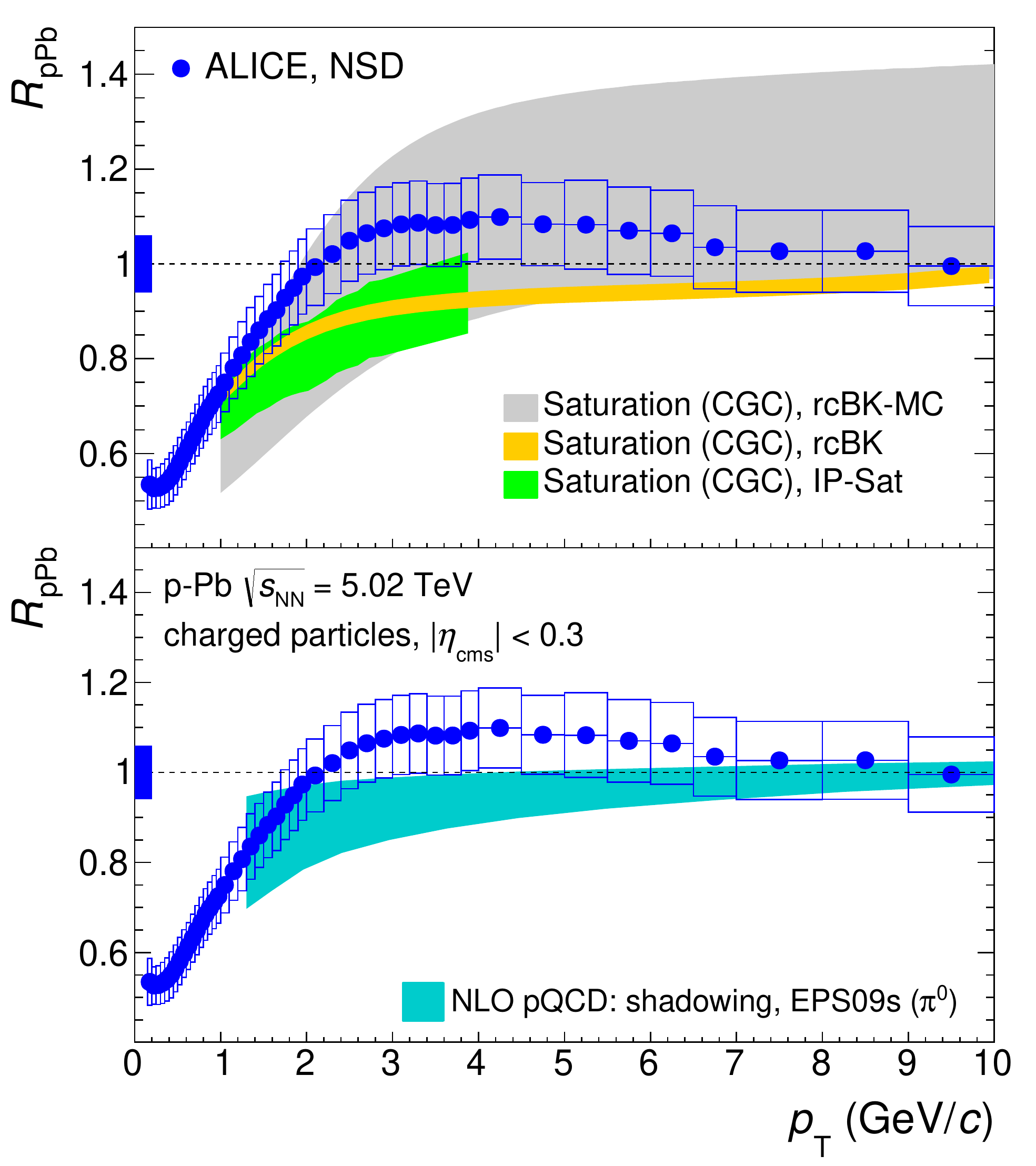}
\ec
\vspace{-.5cm}
\caption{$\RpA^{h}$ measured by {ALICE} compared to theory~\cite{ALICE:2012mj}.}
   \label{fig:h_alice}
\vspace{-.4cm}
\end{figure}
%%%%%%%%%%%%%%%%%%%%%%%%%%%%%%%%%%%%%%%%%%
Parton multiple scattering in a nucleus can be taken into account in the dipole formalism, and more specifically within the color glass condensate (CGC) which allows for the inclusion of non-linear effects in the QCD evolution of a dense nucleus at small $x$~\cite{HP16:CGC}. Several (semi-)hard processes have been investigated in the CGC in pA collisions, yet under different working assumptions in the phenomenological applications. Often used is the `hybrid' formalism according to which the cross section is expressed as a convolution of the usual PDF in the proton (assumed to be in the dilute regime, $Q \gtrsim Q_{s, {\rm p}}$) and the unintegrated gluon distribution in the nucleus ${\cal F}_g^{\rm A}$ (assumed to be in the dense regime, $Q \lesssim Q_{s}$). The latter is computed explicitly in the CGC from the Fourier transform of correlators of Wilson lines (dipoles or quadrupoles) which resum multiple scattering to all orders. Taking the example of light hadron production in pA collisions, the LO cross section can be expressed symbolically as $\sigma^{{\rm pA}\to {\rm h+X}} =  f_{i}^{p}(\mu) \otimes\ {\cal F}_{g}^{A}(\kt) \otimes\ {D}_{i}^h(\mu_{_{\rm F}})$, where $D_i^h$ is the fragmentation function and $\mu_{_{\rm F}}$ the fragmentation scale~\cite{Dumitru:2005gt}. Examples of CGC calculations on light-hadron suppression in pPb collisions at $\sqrts=5.02$~TeV are compared in Fig.~\ref{fig:h_alice} (top panel) to {ALICE} data~\cite{ALICE:2012mj} (results were also presented by ATLAS~\cite{HP16:h_ATLAS}). All calculations predict a common trend, namely an increase of $\RpA$ with $\pt$ due to 
(lesser) shadowing and transverse momentum broadening; however, the magnitude and uncertainty of $\RpA$ differ somehow in the different CGC implementations. While the larger uncertainty of rcBK-MC calculations encompasses {ALICE} results, the other  calculations (rcBK and IP-Sat) stand slightly but systematically below the data.
ALICE results are also compared in Fig.~\ref{fig:h_alice} (bottom) to NLO calculations using EPS09 which prove slightly below the data in the intermediate $\pt$ range, even though the EPS09 individual members sets exhibiting less shadowing would be in reasonable agreement.
The comparison with RHIC measurements~\cite{Adler:2003ii} is very interesting, in particular two aspects which are, perhaps, challenging to understand: (i) the shape and magnitude of $\RpA$ measured at mid-rapidity at RHIC~\cite{Adler:2003ii} and LHC~\cite{ALICE:2012mj,Khachatryan:2016odnAad:2016zif} are very similar, despite a factor of $\sqrt{s_{_{\rm LHC}}} / \sqrt{s_{_{\rm RHIC}}} \simeq 25$ difference in the $x_2$ momentum fraction, (ii) a significantly stronger hadron suppression is observed at RHIC at forward pseudorapidity ($\eta =3.2$) compared to LHC at mid-rapidity, although the value of $x_2$, hence the saturation scale, is similar ($\sqrt{s_{_{\rm LHC}}} / \sqrt{s_{_{\rm RHIC}}}\times  \exp(-3.2) \simeq 1)$. In this respect, it would be informative to revisit RHIC data with up-to-date CGC calculations such as the ones shown in Fig.~\ref{fig:h_alice}; this would give a hint on whether both RHIC and LHC light-hadron data can be understood within a common framework based on saturation.
On a more formal side, a tremendous effort has been performed in order to tackle particle production at NLO in the CGC. Several talks addressed more specifically the issue of negative cross sections reported at large $\pt$~\cite{Stasto:2013cha}, its physical origin and how to cure this artifact~\cite{HP16:CGCNLO}.
%%%%%%%%%%%%%%%%%%%%%%%%%%%%%%%%%%%%%%%%%%
\begin{figure}[tbp]
\bc \includegraphics[width=8.cm]{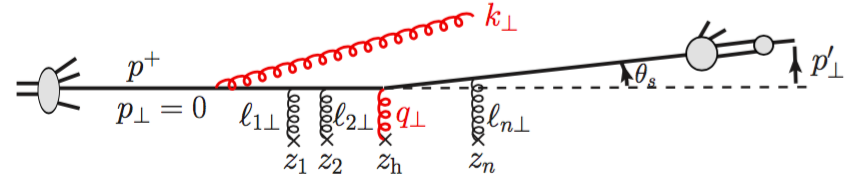}  \ec
\vspace{-.4cm}
\caption{Induced gluon radiation off a fast parton experiencing multiple scattering in a dense nucleus.}
   \label{fig:cartoon_eloss}
\vspace{-.1cm}
\end{figure}
%%%%%%%%%%%%%%%%%%%%%%%%%%%%%%%%%%%%%%%%%%

The fast incoming parton not only scatters elastically in the dense nucleus but could also radiate a gluon which takes part of its energy (see the sketch in Fig.~\ref{fig:cartoon_eloss}). The difference of gluon radiation in a nucleus and in a proton --~ the medium-induced gluon radiation spectrum~-- exhibits different scaling properties depending on the typical gluon formation time $\tf$:
\begin{itemize}
\item[(i)] In the regime of `small' formation time, $\tf \lesssim L$, known as Landau-Pomeranchuk-Migdal (LPM), the mean energy loss scales parametrically as $\Delta E_{_{\rm LPM}} \propto \alpha_s  \hat{q} L^2$, with a dependence on the parton energy at most logarithmic~\cite{Baier:1996skGyulassy:2000fs};
\item[(ii)] Fully coherent energy loss arises from the induced radiation of gluons with formation time much larger than the medium length, $\tf \gg L$~\cite{Arleo:2010rb,Arleo:2012rs}. In this regime, the average energy loss becomes proportional to the parton energy $E$, $\Delta E_{\rm coh} \propto E$, and thus overwhelms (at large $E$) LPM energy loss.
\end{itemize}
At the LHC, the effect of LPM energy loss in nuclear matter, $\Delta E_{_{\rm LPM}} / E \sim 1/E$, should be negligible because the incoming parton energy $E$ (in the nucleus rest frame) is extremely large. The LPM regime would be best probed in semi-inclusive DIS on nuclear targets or Drell-Yan production in pA collisions at lower collision energy; it can also be probed in heavy-ion collisions in which a not too energetic parton propagates in a hot medium. On the contrary, fully coherent energy loss should affect all hard processes even at the LHC as $\Delta E_{\rm coh} / E \sim \alpha_s\times (Q_s/Q)$ is finite in the high-energy limit; an exception concerns hard processes with color neutral final states (such as single production of lepton pairs or massive weak bosons) which are insensitive to coherent energy loss~\cite{Peigne:2014uha}.
%%%%%%%%%%%%%%%%%%%%%%%%%%%%%%%%%%%%%%%%%%
\begin{figure}[tbp]
\bc \includegraphics[width=7.cm]{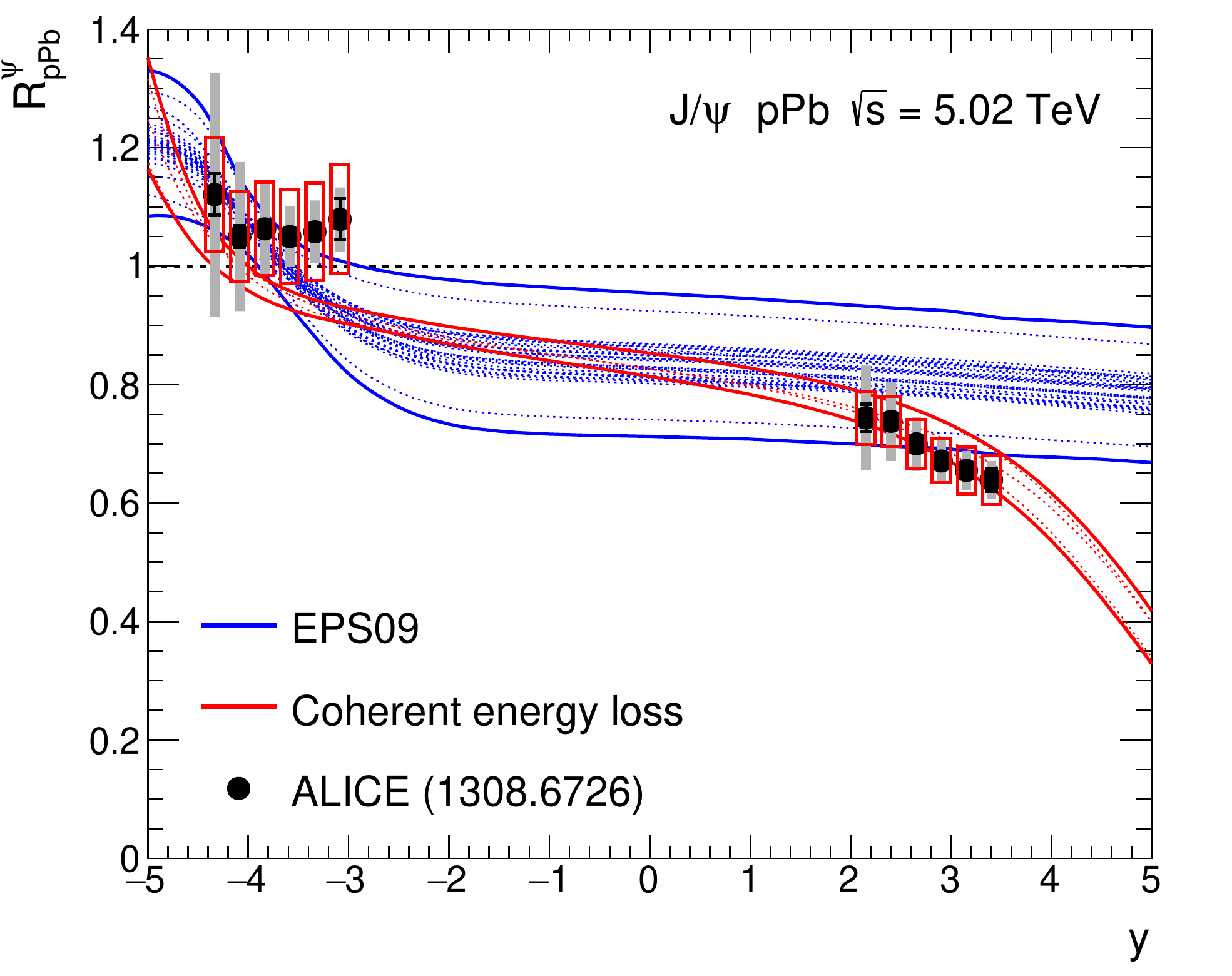}  \ec
\vspace{-.4cm}
\caption{$\RpA^{\jpsi}$ measured by {ALICE} compared to theory.}
   \label{fig:jpsi}
%\vspace{-.2cm}
\end{figure}
%%%%%%%%%%%%%%%%%%%%%%%%%%%%%%%%%%%%%%%%%%

Fully coherent energy loss has been applied successfully to quarkonium ($\jpsi$, $\Upsilon$) suppression in pA collisions, from fixed-target experiments to LHC~\cite{Arleo:2012rs}. In particular, this process allows for a quantitative explanation of $\jpsi$ suppression measured at forward rapidity at all center-of-mass energies. At the LHC, predictions in pPb collisions proved to be in excellent agreement with {ALICE}~\cite{Abelev:2013yxa} (and LHCb~\cite{Aaij:2013zxa}) data, see Fig.~\ref{fig:jpsi}. Updated CGC calculations based on the Color Evaporation Model, presented at this conference, also reproduce very well the experimental results~\cite{Ducloue:2015gfa}. On the contrary, EPS09 NLO calculations based on the sole effects of nPDF stand slightly \emph{above} the data in the most forward rapidity bins\footnote{A simplified (compatible) calculation~\cite{Arleo:2015qiv} is shown in Fig.~\ref{fig:jpsi} as thin dotted lines for the EPS09 member sets and solid lines for the envelope of the uncertainty band.}~\cite{Vogt:2015uba}; interestingly this is opposite  to what is observed in the light hadron production channel (Fig.~\ref{fig:h_alice}). Similarly, the $D$ meson forward-over-backward production ratio reported by LHCb during the conference~\cite{HP16:D_LHCb} turns out to be slightly lower than (yet compatible with) EPS09 NLO calculations; the calculation of coherent energy loss effects of open heavy-flavour $\RpA$ is in progress.
Hopefully, future pPb data at $\sqrt{s}=5.02$~TeV (using actual pp \emph{data} at this energy) and at $\sqrt{s}=8.16$~TeV will allow one to clarify the origin of quarkonium suppression at the LHC. Drell-Yan measurements in pPb collisions at LHC may also play a key role in clarifying the respective effects of nPDF and coherent energy loss~\cite{Arleo:2015qiv}.
Another interesting observation is the stronger suppression suffered by quarkonium excited states in pPb collisions, $\RpA^{\psi({\rm 2S})} < \RpA^{\psi({\rm 1S})}$~\cite{Chatrchyan:2013nza}. Since nPDF effects and coherent energy loss effects are expected to suppress equally 1S and 2S states, it is possible that the particles produced in pPb collisions dissociate excited states which are more loosely bound~\cite{Ferreiro:2014bia}.
%%%%%%%%%%%%%%%%%%%%%%%%%%%%%%%%%%%%%%%%%
\section{Beyond the Glauber model}
\label{sec:hardsoft}
%%%%%%%%%%%%%%%%%%%%%%%%%%%%%%%%%%%%%%%%%

Obviously of interest is the medium length dependence of the different effects mentioned above. It is often investigated differently at fixed-target and at collider experiments:
\bi
\item[(i)] Fixed-target facilities allow for measuring hard processes in minimum bias pA collisions on different nuclear targets;
\item[(ii)] At colliders, changing nuclei is of course more challenging. Instead, pA collisions on a single nucleus are binned in terms of `event activity' (either multiplicity or energy distributions) which may be correlated to the centrality of the collision.
\ei
Comparing different nuclear targets appears, at least to me, more satisfactory. An obvious advantage is the definition of $\RpA$, which does not depend on quantities like thickness functions to be determined in a Glauber model and which comes along with uncertainties. On the contrary, the correlation between event activity and collision centrality in pPb collisions at the LHC appears rather loose, unlike that in heavy-ion collisions, making more delicate the interpretation of the data since a given event activity would correspond to a wide range of centralities. Another aspect may be problematic. So far we have discussed the production of  {\it inclusive} observables, pA$\to$~(hard~particle)~+~X; this is an important requirement for the collinear factorization theorem to apply. When looking at the rate of a given hard process as a function of the event activity, the final state is different and less inclusive, pA$\to$~(hard~particle~+ specific~activity)~+~X. Computing this final state is arduous as it requires to model not only the soft underlying dynamics but also its correlation with specific hard processes. Moreover, the interpretation of $\RpA$ becomes dubious since the two processes which are compared, in pp and in pA collisions, are different by definition. 

When the phase space to produce a hard process is restricted, typically when the projectile or target momentum fraction is large, $x_{1,2} \sim Q/\sqrt{s} \lesssim 1$, the hard process and the event activity may no longer `factorize' (in other words, there's no such thing as an `underlying' activity) since less energy is available to produce the event activity; as a consequence such hard processes are likely to be labeled as a `peripheral' pA collision independent of the actual collision centrality~\cite{Perepelitsa:2014yta,Armesto:2015kwa}.
An example of such a `event bin migration' could be found in the $\RpA$ measurement of large $\pt$ jets by ATLAS~\cite{ATLAS:2014cpa}: data exhibit a significant depletion of jets at large $\pt$ in more `central' events, and an enhancement in more `peripheral' events. Although it is not possible to conclude firmly on the origin of these observations, it is interesting to note that when no selection is made on the event activity, hence for minimum bias collisions, the measured value of $\RpA$ is consistent with unity, which seems to corroborate the bias due to specific event activity; it is also confirmed by Monte Carlo studies~\cite{Armesto:2015kwa}. Similar observations have also been found by PHENIX at RHIC~\cite{Adare:2015gla}. Although they complicate somehow the interpretation of hard processes in pA collisions, such correlations are interesting on their own as they might reveal interesting aspects of particle production dynamics~\cite{Perepelitsa:2014yta,Armesto:2015kwa,Alvioli:2014edaKordell:2016njgMcGlinchey:2016ssj} which understanding may help to constrain Monte Carlo event generators.

\section*{Acknowledgements}
It is a pleasure to thank R. G\'en\'eral de Cassagnac, M. Nguyen, and S. Peign\'e for useful comments on the manuscript, and A. Andronic for providing Fig.~\ref{fig:h_alice}.

%\bibliographystyle{mybst_noexp}
%\bibliography{mybib}

\begin{thebibliography}{10}

\bibitem{Eskola:2009uj}
{\bf EPS09}, K.~J. Eskola, H.~Paukkunen and C.~A. Salgado,  JHEP {\bf 04} (2009) 065
  [\href{http://arXiv.org/abs/0902.4154}{{\tt 0902.4154}}],
{\bf DSSZ}, D.~de~Florian, R.~Sassot, P.~Zurita and M.~Stratmann,  Phys.\ Rev. {\bf D85}
  (2012) 074028 [\href{http://arXiv.org/abs/1112.6324}{{\tt 1112.6324}}],
{\bf nCTEQ15}, K.~Kovarik {\it et~al.},  Phys. Rev. {\bf D93} (2016) 085037
  [\href{http://arXiv.org/abs/1509.00792}{{\tt 1509.00792}}],
{\bf EPPS16}, K.~J. Eskola, P.~Paakkinen, H.~Paukkunen and C.~A. Salgado,
  \href{http://arXiv.org/abs/1612.05741}{{\tt 1612.05741}},
{\bf KA15}, H.~Khanpour and S.~Atashbar~Tehrani,  Phys. Rev. {\bf D93} (2016) 014026
  [\href{http://arXiv.org/abs/1601.00939}{{\tt 1601.00939}}].

\bibitem{Helenius:2016hcu}
I.~Helenius, H.~Paukkunen and N.~Armesto,  PoS {\bf DIS2016} (2016) 276
  [\href{http://arXiv.org/abs/1606.09003}{{\tt 1606.09003}}].

\bibitem{Paukkunen:2014zia}
H.~Paukkunen and P.~Zurita,  JHEP {\bf 12} (2014) 100
  [\href{http://arXiv.org/abs/1402.6623}{{\tt 1402.6623}}].

\bibitem{Armesto:2015lrg}
N.~Armesto, H.~Paukkunen, J.~M. Pen{\'\i}n, C.~A. Salgado and P.~Zurita,  Eur.
  Phys. J. {\bf C76} (2016) 218 [\href{http://arXiv.org/abs/1512.01528}{{\tt
  1512.01528}}].

\bibitem{Paukkunen:2014pha}
H.~Paukkunen, K.~J. Eskola and C.~Salgado,  Nucl. Phys. {\bf A931} (2014) 331
  [\href{http://arXiv.org/abs/1408.4563}{{\tt 1408.4563}}].

\bibitem{Paukkunen:2010qg}
H.~Paukkunen and C.~A. Salgado,  JHEP {\bf 1103} (2011) 071
  [\href{http://arXiv.org/abs/1010.5392}{{\tt 1010.5392}}].

\bibitem{Arleo:2015qiv}
F.~Arleo and S.~Peign{\'e},  \href{http://arXiv.org/abs/1512.01794}{{\tt
  1512.01794}}, to appear in Phys. Rev. D

\bibitem{Arleo:2007js}
F.~Arleo and T.~Gousset,  Phys. Lett. {\bf B660} (2008) 181
  [\href{http://arXiv.org/abs/0707.2944}{{\tt 0707.2944}}],
F.~Arleo, K.~J. Eskola, H.~Paukkunen and C.~A. Salgado,  JHEP {\bf 04} (2011)
  055 [\href{http://arXiv.org/abs/1103.1471}{{\tt 1103.1471}}].

\bibitem{dEnterria:2015mgr}
D.~d'Enterria, K.~Krajcz{\'a}r and H.~Paukkunen,  Phys. Lett. {\bf B746} (2015)
  64 [\href{http://arXiv.org/abs/1501.05879}{{\tt 1501.05879}}].

\bibitem{CMS:2016kjd}
{\bf CMS}, PAS-HIN-16-003 and Y.-J. Lee, these proceedings

\bibitem{ATLAS:2016ckg}
{\bf ATLAS}, Phys. Rev. {\bf C92} (2015),
  044915 [\href{http://arXiv.org/abs/1507.06232}{{\tt 1507.06232}}], 
and M.~Dumancic, these proceedings.

\bibitem{Khachatryan:2015hha}
{\bf CMS}, Phys. Lett. {\bf B750} (2015) 565
  [\href{http://arXiv.org/abs/1503.05825}{{\tt 1503.05825}}].

\bibitem{Ru:2016ifu}
P.~Ru and B.-W. Zhang,   \href{http://arXiv.org/abs/1612.02899}{{\tt 1612.02899}}, these proceedings.

\bibitem{Arleo:2015dba}
F.~Arleo, {\'E}.~Chapon and H.~Paukkunen,  Eur. Phys. J. {\bf C76} (2016) 214
  [\href{http://arXiv.org/abs/1509.03993}{{\tt 1509.03993}}].

\bibitem{Catani:2009sm}
S.~Catani, L.~Cieri, G.~Ferrera, D.~de~Florian and M.~Grazzini,  Phys. Rev.
  Lett. {\bf 103} (2009) 082001 [\href{http://arXiv.org/abs/0903.2120}{{\tt
  0903.2120}}].

\bibitem{Graziani:2145943}
{\bf LHCb}, G.~Graziani, {\it el al.},  Tech. Rep. LHCb-PUB-2016-011.

\bibitem{HP16:CGC}
R.~Venugopalan, these proceedings.

\bibitem{Dumitru:2005gt}
A.~Dumitru, A.~Hayashigaki and J.~Jalilian-Marian,  Nucl. Phys. {\bf A765}
  (2006) 464 [\href{http://arXiv.org/abs/hep-ph/0506308}{{\tt
  hep-ph/0506308}}].

\bibitem{ALICE:2012mj}
{\bf {ALICE}},  Eur.  Phys. J. {\bf C74} (2014) 3054 [\href{http://arXiv.org/abs/1405.2737}{{\tt
  1405.2737}}].

\bibitem{HP16:h_ATLAS}
{\bf ATLAS}, P.~Balek, these proceedings.

\bibitem{Adler:2003ii}
{\bf PHENIX}, Phys. Rev. Lett. {\bf 91} (2003) 072303
  [\href{http://arXiv.org/abs/nucl-ex/0306021}{{\tt nucl-ex/0306021}}], 
{\bf BRAHMS}, Phys. Rev. Lett. {\bf 93} (2004) 242303
  [\href{http://arXiv.org/abs/nucl-ex/0403005}{{\tt nucl-ex/0403005}}].

\bibitem{Khachatryan:2016odnAad:2016zif}
{\bf CMS}, \href{http://arXiv.org/abs/1611.01664}{{\tt 1611.01664}},
{\bf ATLAS}, Phys. Lett. {\bf B763} (2016) 313
  [\href{http://arXiv.org/abs/1605.06436}{{\tt 1605.06436}}].

\bibitem{Stasto:2013cha}
A.~M. Sta{\'s}to, B.-W. Xiao and D.~Zaslavsky,  Phys. Rev. Lett. {\bf 112} (2014)
  012302 [\href{http://arXiv.org/abs/1307.4057}{{\tt 1307.4057}}].

\bibitem{HP16:CGCNLO}
G.~Beuf, T.~Lappi and Y.~Zhu, these proceedings.

\bibitem{Baier:1996skGyulassy:2000fs}
R.~Baier,  {\it et~al.},  Nucl.
  Phys. {\bf B484} (1997) 265 [\href{http://arXiv.org/abs/hep-ph/9608322}{{\tt
  hep-ph/9608322}}], 
M.~Gyulassy, P.~L{\'e}vai and I.~Vitev,  Phys. Rev. Lett. {\bf 85} (2000) 5535
  [\href{http://arXiv.org/abs/nucl-th/0005032}{{\tt nucl-th/0005032}}].

\bibitem{Arleo:2010rb}
F.~Arleo, S.~Peign{\'e} and T.~Sami,  Phys. Rev. {\bf D83} (2011) 114036
  [\href{http://arXiv.org/abs/1006.0818}{{\tt 1006.0818}}].

\bibitem{Arleo:2012rs}
F.~Arleo and S.~Peign{\'e},  JHEP {\bf 03} (2013) 122
  [\href{http://arXiv.org/abs/1212.0434}{{\tt 1212.0434}}], F.~Arleo, R. Kolevatov, M. Rustamova, S. Peign{\'e}, JHEP {\bf 05} (2013) 155 [\href{http://arXiv.org/abs/1304.0901}{{\tt 1304.0901}}].

\bibitem{Peigne:2014uha}
F.~Arleo, R.~Kolevatov and S.~Peign{\'e},  Phys. Rev. {\bf D93} (2016) 014006
  [\href{http://arXiv.org/abs/1402.1671}{{\tt 1402.1671}}],
S.~Peign{\'e} and R.~Kolevatov,  JHEP {\bf 01} (2015) 141
  [\href{http://arXiv.org/abs/1405.4241}{{\tt 1405.4241}}].

\bibitem{Abelev:2013yxa}
{\bf {ALICE}}, JHEP {\bf 1402} (2014) 073
  [\href{http://arXiv.org/abs/1308.6726}{{\tt 1308.6726}}].

\bibitem{Aaij:2013zxa}
{\bf LHCb}, JHEP {\bf 1402} (2014) 072
  [\href{http://arXiv.org/abs/1308.6729}{{\tt 1308.6729}}].

\bibitem{Ducloue:2015gfa}
B.~Duclou{\'e}, T.~Lappi and H.~M{\"a}ntysaari,  Phys. Rev. {\bf D91} (2015)
  114005 [\href{http://arXiv.org/abs/1503.02789}{{\tt 1503.02789}}].

\bibitem{Vogt:2015uba}
R.~Vogt,  Phys. Rev. {\bf C92} (2015) 034909
  [\href{http://arXiv.org/abs/1507.04418}{{\tt 1507.04418}}].

\bibitem{HP16:D_LHCb}
{\bf LHCb}, LHCb-CONF-2016-003 and X. Zhu, these proceedings.% .

\bibitem{Chatrchyan:2013nza}
{\bf CMS}, JHEP {\bf 1404} (2014) 103
  [\href{http://arXiv.org/abs/1312.6300}{{\tt 1312.6300}}], 
{\bf {ALICE}}, JHEP {\bf 12} (2014) 073
  [\href{http://arXiv.org/abs/1405.3796}{{\tt 1405.3796}}], 
{\bf LHCb}, JHEP {\bf 03} (2016) 133
  [\href{http://arXiv.org/abs/1601.07878}{{\tt 1601.07878}}].

\bibitem{Ferreiro:2014bia}
E.~G. Ferreiro,  Phys. Lett. {\bf B749} (2015) 98 
  [\href{http://arXiv.org/abs/1411.0549}{{\tt 1411.0549}}], J.-P. Lansberg, these proceedings.

\bibitem{Perepelitsa:2014yta}
D.~V. Perepelitsa and P.~A. Steinberg,
  \href{http://arXiv.org/abs/1412.0976}{{\tt 1412.0976}}.

\bibitem{Armesto:2015kwa}
N.~Armesto, D.~C. G{\"u}lhan and J.~G. Milhano,  Phys. Lett. {\bf B747} (2015)
  441 [\href{http://arXiv.org/abs/1502.02986}{{\tt 1502.02986}}].

\bibitem{ATLAS:2014cpa}
{\bf ATLAS}, Phys. Lett. {\bf B748} (2015) 392
  [\href{http://arXiv.org/abs/1412.4092}{{\tt 1412.4092}}].

\bibitem{Adare:2015gla}
{\bf PHENIX}, Phys. Rev. Lett. {\bf 116} (2016) 122301
  [\href{http://arXiv.org/abs/1509.04657}{{\tt 1509.04657}}].

\bibitem{Alvioli:2014edaKordell:2016njgMcGlinchey:2016ssj}
M.~Alvioli {\it et al.},  Phys.
  Rev. {\bf C93} (2016) 011902 [\href{http://arXiv.org/abs/1409.7381}{{\tt
  1409.7381}}],
M.~Kordell and A.~Majumder,  \href{http://arXiv.org/abs/1601.02595}{{\tt
  1601.02595}},
D.~McGlinchey, J.~L. Nagle and D.~V. Perepelitsa,  Phys. Rev. {\bf C94} (2016)
  024915 [\href{http://arXiv.org/abs/1603.06607}{{\tt 1603.06607}}].

\end{thebibliography}

\providecommand{\href}[2]{#2}\begingroup\raggedright\endgroup

\end{document}